\documentclass{article}

\usepackage[dvipsnames]{xcolor}
\usepackage{microtype}
\usepackage{graphicx}
\usepackage{subfigure}
\usepackage{booktabs} 
\usepackage{comment}

\usepackage{hyperref}

\usepackage[accepted]{icml2024}

\usepackage{amsmath}
\usepackage{amssymb}
\usepackage{mathtools}
\usepackage{amsthm}

\usepackage[capitalize,noabbrev]{cleveref}

\theoremstyle{plain}

\theoremstyle{definition}

\theoremstyle{remark}

\usepackage[textsize=tiny]{todonotes}
\newcommand{\rebuttal}[1]{#1}

\icmltitlerunning{Why Do Animals Need Shaping?}
\allowdisplaybreaks

\begin{document}

\twocolumn[
\icmltitle{Why Do Animals Need Shaping? \\ A Theory of Task Composition and Curriculum Learning}



\icmlsetsymbol{equal}{*}

\begin{icmlauthorlist}

\icmlauthor{Jin Hwa Lee}{ucl}
\icmlauthor{Stefano Sarao Mannelli}{ucl}
\icmlauthor{Andrew Saxe}{ucl}
\end{icmlauthorlist}
\icmlaffiliation{ucl}{Sainsbury Wellcome Centre for Neural Circuits and Behaviour \& Gatsby Computational Neuroscience Unit, University College London, London, UK\\}
\icmlcorrespondingauthor{Jin Hwa Lee}{jin.lee.22@ucl.ac.uk}


\vskip 0.3in
]

\printAffiliationsAndNotice{} 

\begin{abstract}
Diverse studies in systems neuroscience begin with extended periods of \rebuttal{curriculum} training known as `shaping' procedures. These involve progressively studying component parts of more complex tasks, and can make the difference between learning a task quickly, slowly or not at all. Despite the importance of shaping to the acquisition of complex tasks, there is as yet no theory that can help guide the design of shaping procedures, or more fundamentally, provide insight into its key role in learning. Modern deep reinforcement learning systems might implicitly learn compositional primitives within their multilayer policy networks. Inspired by these models, we propose and analyse a model of deep policy gradient learning of simple compositional reinforcement learning tasks. Using the tools of statistical physics, we solve for exact learning dynamics and characterise different learning strategies including \textit{primitives pre-training}, in which task primitives are studied individually before learning compositional tasks. We find a complex interplay between task complexity and the efficacy of shaping strategies. Overall, our theory provides an analytical understanding of the benefits of shaping in a class of compositional tasks and a quantitative account of how training protocols can disclose useful task primitives, ultimately yielding faster and more robust learning.
\end{abstract}

\section{Introduction}
\label{Introduction}
Shaping is critical for effective learning in animals and humans~\cite{skinner2019behavior,pavlov1927conditioned}. Rather than teaching a complex task directly, shaping aims to gradually teach components of the complex task. Imagine a mouse in a laboratory learning a decision-making task to get water rewards. To learn this task, the mouse needs to learn many sub-tasks often through reinforcement, for example, how to interact with a reward port, or how a light bulb predicts rewards. Depending on what shaping procedures are applied, the animals might learn faster, slower or even completely fail to learn a given task. Indeed, many studies in neuroscience exploit shaping stages for behavioural training of animals~\cite{mushiake2001visually,laboratory2021standardized,grossman2022serotonin,makino2023arithmetic}. In humans, studies similarly show that curriculum---the order in which tasks are trained---is important for generalisation~\cite{elio1984effects,clerkin2017real,pashler2013does,eckstein2021mind,dekker2022curriculum}. Nevertheless, we do not have a theory that can quantitatively explain the role of shaping and how it changes the learning dynamics of intelligent systems which could give us deeper insights into these procedures and guide design choices in experiments.

One crucial feature of shaping is compositionality. Shaping stages involve breaking down complex tasks into simpler components and learning them individually before animal learns to integrate those primitive tasks in the right context to complete the higher level tasks~\cite{schulz2017compositional,hupkes2020compositionality}. In other words, shaping is a form of \textit{curriculum learning}, leveraging the compositional structure of complex tasks. Many real world tasks require mastering different skills and combining them in the right way. This property is often referred to as systematic compositionality, which enables us to flexibly reuse previously acquired \textit{primitives} by combining them~\cite{chomsky2014aspects,smolensky1990tensor,DBLP:conf/cogsci/LakeLB19,dehaene2022symbols}.

On the other hand, the recent success of deep reinforcement learning (RL) in highly complex tasks motivates using these models to study compositional task and curriculum learning. Many empirical studies indicate that multilayer policy networks might implicitly learn compositional primitives if not explicitly~\cite{Baker2020Emergent,zhang2020automatic,nachum2018data,andreas2017modular,barreto2020fast}. This motivates using a deep policy network model of compositional RL tasks to study learning dynamics and understand the role of shaping. 

Our goal in this paper is to develop a simple theory of compositional task learning, to obtain conceptual insight into the factors affecting learning performance. The theoretical formulation of compositional learning in RL is challenging due to the limited mathematical tools available. In this study, we borrow tools from statistical mechanics and the recent framework of analyzing RL dynamics using the teacher-student setup~\cite{patel2023rl,bordelon2023loss} to shed light on the learning dynamics of compositional learning in RL. We characterize different curricula including \textit{primitives pre-training} and \textit{vanilla training}. We reveal that different curricula result in differences in training time and robustness to noise during training, indicating potential ramifications of shaping and curriculum learning in complex compositional tasks.

\subsection*{Main contributions}
\begin{itemize}
    \item We model a compositional RL task and its primitives in the teacher-student setup and derive an asymptotically exact set of ordinary differential equations (ODEs) describing the evolution of the order parameters in online learning. 
    \item We analyze learning dynamics under two different curricula: 1) \textit{primitives pre-training}, in which an agent is first trained on primitives before learning the compositional task; and 2) \textit{vanilla training}, in which an agent is trained directly on the compositional task.
    \item We investigate the benefits of curriculum learning based on our learning dynamics analysis---principally, faster learning and robustness to noise. 
    \item Finally, we connect our study to the literature from psychology and neuroscience. 
\end{itemize}

\subsection*{Related Work}

The \textbf{teacher-student framework} and the tools of statistical physics have been widely used to study learning dynamics in neural network models~\cite{gardner1989three, seung1992statistical, biehl1995learning, saad1995exact,engel2001statistical, zdeborova2016statistical}. Diverse learning behaviors and phenomena in different settings have been studied using the paradigm~\cite{goldt2019dynamics,mannelli2019passed,advani2020high,bahri2020statistical,saglietti2022analytical,gerace2022probing}. In particular, our model formulation of a compositional task is largely inspired by the RL perceptron~\cite{patel2023rl}, a recent theoretical framework to study the average dynamics of a prototypical policy gradient RL network in high-dimensions. The agent faces a high-dimensional input and learns to select actions through sparse rewards. In our work, we introduce hierarchy to the RL perceptron to bring a notion of primitives and higher-level task composition, which is defined by a combination of primitives.

\textbf{Curriculum learning} has been studied extensively in neuroscience and psychology \cite{ahissar1997task,hornsby2014improved,skinner2019behavior,dekker2022curriculum, makino2023arithmetic, hocker2024curriculum} across different species (humans, rodents, pigeons) and tasks (decision making, perceptual learning, motor learning).
In the machine learning community, while its importance has long been acknowledged~\cite{bengio2009curriculum,wang2021survey}, a recent survey shows only marginal benefits of curriculum in standard supervised learning benchmarks~\cite{wu2021when}. \rebuttal{However, the right curricula have shown greater effects in the context of reinforcement learning~\cite{narvekar2020curriculum, tessler2017deep,karpathy2012curriculum,tong2023adaptive}, especially with hierarchical task structure and network architectures.}

Leveraging \textbf{compositionality} for efficient machine learning systems has been proposed in many modalities and contexts. In vision models, disentangled representations seek to factorize input variance into semantic constituents, modeling the input data as a composition of those factors~\cite{montero2020role}. In the domain of language, meta-training on memory-augmented neural networks exhibits certain forms of compositional generalisation~\cite{lake2019compositional}. \rebuttal{Furthermore, it has been shown that modular motifs emerge from training with multiple cognitive tasks and they can be repurposed to rapidly learn new tasks that combine these motifs~\cite{driscoll2022flexible,yang2019task}.} Yet theoretical understanding of learning compositional tasks is limited, though there have been recent studies which formulate compositionality in a data-generating process perspective~\cite{wiedemer2024compositional, okawa2023compositional}. 

In RL, compositionality has been often studied in the context of temporally extended action planning. Hierarchical RL instantiates the idea of decomposing planning using temporal abstraction to chain learned skills to perform a complex behavior on a longer timescale~\cite{sutton1999between,bacon2017option}. The compositionality in our study is conceptually more similar to \textit{functional composition}, which has been addressed with explicitly modularised neural networks for each task or policy primitive~\cite{devin2017learning,saxe2017hierarchy, goyal2021recurrent, yang2020multi,mittal2020learning, mendez2022modular}. In a similar spirit, recent theoretical work showed that modular networks can discover compositional structure through learning sparse combinations of modules~\cite{schug2024discovering}. Our work focuses on compositional learning in RL with policy gradient methods.

\begin{figure}[hbt!]
\vskip 0.2in
\begin{center}
\centerline{\includegraphics[width=\columnwidth]{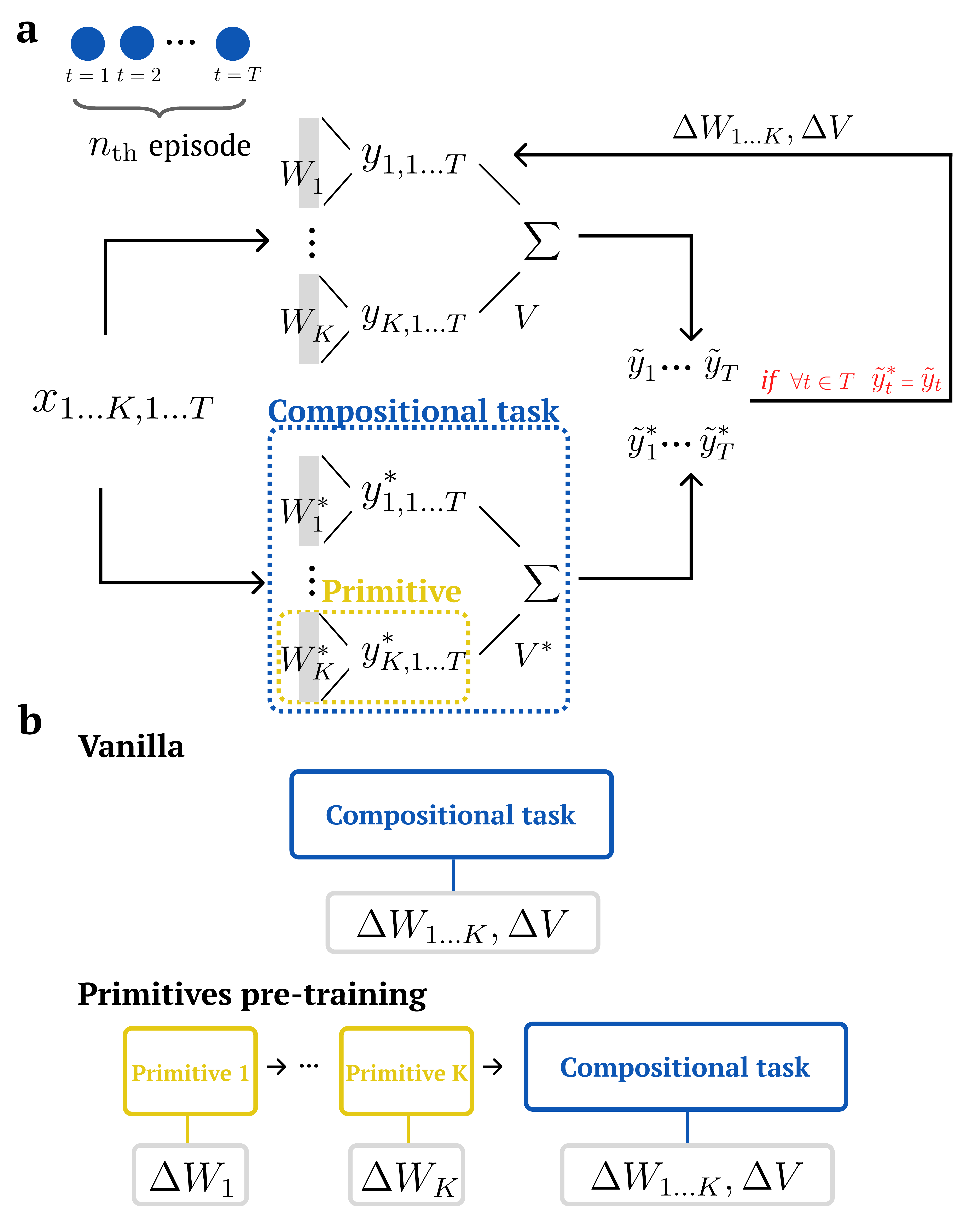}}
\caption{\textbf{Schema of primitives and a compositional task in the teacher-student setup.} \textbf{a)} Illustration of a \textit{compositional task} with $K$ \textit{primitives} in the teacher-student setup. Each primitive $k=1...K$ is modelled by a fixed teacher network $W^*_k$ and learnable student network $W_k$. At each timestep $t$ of an episode with total length $T$, the output of each primitive teacher $W^*_k$ from task-dependent input $x_{k,t}$ is blended using fixed teacher context $V^*_k$ to generate a compositional choice $\Tilde{y}^*_t$. The goal of learning for the student primitives $W_k$ and context $V$ is to generate the same compositional choice $\Tilde{y}_t$ as the teacher choice. Importantly, the student updates the weights only when it makes all $T$ choices in the episode correctly, reflecting a sparse reward setting. \textbf{b)} Schematic of two curricula: vanilla training and primitives pre-train
ing. In vanilla training, an agent directly learns the compositional task. In primitives pre-training, an agent learns each primitive first to expert level and then learns the compositional task.}
\label{figure1}
\end{center}
\vskip -0.3in
\end{figure}

\section{Compositional Learning and Primitives}
\rebuttal{To begin with, we conceptualize compositional learning and primitives with the following example. Ito et al. (\citeyear{ito2022compositional}) studied a compositional task that consists of a combination of rules of three domains (sensory, logic, motor) that a subject needs to combine to make a decision. For example, the sensory rule can be ``red" or ``blue", the logic rule can be ``both" or ``either", and the motor rule can be ``right" or ``left". With such rules, an example compositional task built from the rules could be ``If $<\textbf{both}>$ stimuli are $<\textbf{red}>$, press $<\textbf{right}>$". The primitives in this context are individual rules, which need to be selected and combined appropriately to make a correct decision for the compositional task.} \rebuttal{Following this concept,} we formulate a compositional generalisation task with \textit{primitives} and \textit{compositional context}, where a compositional task is modeled as a linear combination of the primitives given by the context. 

\rebuttal{In order to theoretically evaluate a network trained to perform a given task, we need a tractable data generative model that generates the ground-truth input-output mapping that a network needs to learn. One standard way is using the teacher-student framework~\cite{seung1992statistical, zdeborova2016statistical}. Within this framework, a teacher is a network used to label randomly sampled inputs, while a student is a network whose goal is to optimise its weights and match the label generated by the teacher. The teacher network can be thought of as an oracle network or an ideal post-learning weight configuration.} Using a teacher-student setup as a generative model, we construct synthetic inputs $x$ and labels $y^*$ from a teacher network (we will denote the teacher network and correct labels with $*$), as schematized in Figure~\ref{figure1}a.

In the following, we describe our teacher-student model of \textit{primitives} and \textit{compositional tasks}.

\paragraph{Primitives.} 
We define $K$ primitive tasks, where each primitive $k=1,...,K$ is constructed with an independent teacher-student pair. Specifically, each primitive task is represented as an RL perceptron~\cite{patel2023rl} which allows us to model policy learning with high-dimensional inputs. In each primitive task, the student needs to make $T$ binary choices: $T$ will be an important parameter of our problem acting as a proxy for difficulty that we call \textit{episode length}. At each time step $t$ of primitive task $k$, the student makes a binary choice given a task-dependent high-dimensional input $x_{k,t}\sim\mathcal{N}(\pmb 0, \mathbb{I}^N)$ of dimension $N$. 
At time step $t$, the choice of student is correct if $y_{k,t}=\text{sign}(W_k \cdot x_{k,t}/\sqrt{N})$ matches $y^*_{k,t}=\text{sign}(W^*_k \cdot x_{k,t}/\sqrt{N})$ provided by the teacher. 
After the student makes choices for $t=1,...,T$, the student receives a delayed reward and updates its weights following a perceptron-like update rule,
\begin{equation}
    \Delta W_k^{n+1} = \frac{\eta_{ w}}{\sqrt{N}}\Biggl(\frac{1}{T}\sum^T_{t=1}y_{k,t} x_{k,t}\Biggr) \mathbb{I}(\Phi(y_{k, 1:T}, y^*_{k, 1:T}))
    \label{eq:primitive_update}
\end{equation}
where $n$ indicates the episode index, $\mathbb{I}$ is an indicator function and $\Phi$ is a boolean criterion (i.e., the student receives a sparse reward and updates its weights at the end of each episode only when it makes all $T$ choices correctly). Figure~\ref{figure1}a illustrates the weight update scheme. \rebuttal{We use this specific learning rule since \citet{patel2023rl} showed that this is analogous to the REINFORCE algorithm~\cite{sutton1999policy}, a widely adopted policy gradient learning algorithm. See Appendix~\ref{appendix:rl_perceptron} for details.}

\paragraph{Compositional Task.}
The compositional task is constructed by blending the outputs of $K$ primitives via a context vector $V^* \in \mathbb{R}_{+}^K$. We denote variables related to the compositional task with `` $\Tilde{~}$ ''. The correct choice at timestep $t$ in the compositional task is given by $\tilde{y}_t^* = \text{sign}\left(\sum^K_{k=1} V^*_k \frac{W^*_k \cdot x_{k,t}}{\sqrt{N}}\right)$, a linear combination of teacher primitives and context weights. Similarly, the student makes a choice from student primitives and context weights, $\tilde y_t = \text{sign}\left(\sum^K_{k=1} V_k \frac{W_k \cdot x_{k,t}}{\sqrt{N}}\right)$. This can be interpreted as linearly combining a set of task rules in an appropriate context to generate the choice. The scheme is shown in the blue-dotted rectangle of Figure~\ref{figure1}a. 

\rebuttal{We derive the compositional version of the Eq.~\ref{eq:primitive_update}, which allows us to model a policy gradient learning equivalent to REINFORCE on this compositional task.} 
When the student is trained for the compositional task, both primitives and context are updated by the following update rules:
\begin{equation}
    \Delta W_k^{n+1} = \frac{\eta_{ w}}{\sqrt{N}}\Biggl(\frac{1}{T}\sum^{T}_{t=1}\Tilde{y}_t V_kx_{k,t}\Biggr) \mathbb{I}(\Phi(\Tilde{y}_{1:T}, \Tilde{y}^*_{1:T})),
    \label{eq:compositional_w_update}
\end{equation}
\begin{equation}
    \Delta V_k^{n+1} = \frac{\eta_{ v}}N\Biggl(\frac{1}{T}\sum^{T}_{t=1}\Tilde{y}_t \frac{W_k \cdot x_{k,t}}{\sqrt{N}}\Biggr)\mathbb{I}(\Phi(\Tilde{y}_{1:T}, \Tilde{y}^*_{1:T})),
    \label{eq:compositional_v_update}
\end{equation}
where $\eta_w$ and $\eta_v$ are learning rates for the primitive weights and context weights, respectively. Analogously to the primitive task update rule, the student only updates its primitive weights and context weights at the end of each episode, when all $T$ choices are made correctly. This learning rule is a proxy of policy gradient learning for our model. \rebuttal{See Appendix~\ref{appendix:rl_perceptron} for details.} 

\rebuttal{Note that our model of compositional tasks relies on learning each primitive independently using separate networks and combining them post hoc, which is only one form of compositional learning.
This is equivalent to assume that primitives are orthogonal which naturally yields a modular structure similar to that in the literature~\cite{schug2024discovering,jarvis2023specialization,driscoll2022flexible}.
}

\section{Learning Dynamics in Different Curricula}

\begin{figure*}[ht]
\vskip 0.2in
\begin{center}
\centerline{\includegraphics[width=\textwidth]{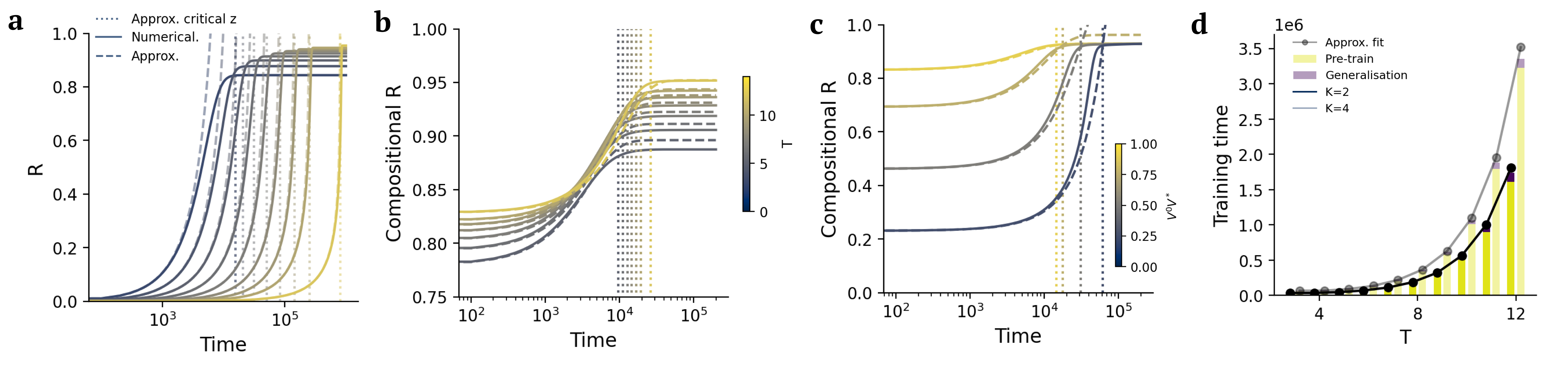}}
\caption{\textbf{Learning dynamics for primitives pre-training curriculum.} \textbf{a)} Learning dynamics of a single primitive task of episode length $T$. The dashed line depicts the approximate dynamics of $R$ from Eq.~\ref{eq:comp_gen_rk}, which captures the numerical integration of the ODE near the beginning of learning ($N=1000, \eta_w = 1$). The dotted line denotes the approximate timescale $\tau$.~\textbf{b)} Learning dynamics during the compositional generalization phase after primitives pre-training across different $T$ ($N=1000, \eta_w = 1, \eta_v = 1, K=2, V^0V^* = 0.5$). \textbf{c)} Learning dynamics of compositional generalization across different cosine similarities between initial $V^0$ and target $V^*$ ($N=1000, \eta_w = 1, \eta_v = 1, K=2, T=8$). 
 \textbf{d)} \rebuttal{Simulated total training time (bars) and approximated total training time from our analytical derivation (lines) for the primitives pre-training curriculum across different $T$ (x-axis) and $K$ (indicated by different transparency). The pre-training time of the primitives (yellow) is the major burden compared to the relatively faster compositional generalisation training time (purple) ($\eta_w = 1, \eta_v = 1, V^0V^* = 0.5$).}}
\label{figure2}
\end{center}
\vskip -0.3in
\end{figure*}
Often curriculum learning in machine learning~\cite{bengio2009curriculum,wu2021when,saglietti2022analytical} is implemented by feeding training data in order of difficulty. The concept of compositionality adds a degree of complexity in task structure and naturally suggests a curriculum for primitives and compositional tasks. In this section, we introduce two curriculum strategies---\textit{primitives pre-training} and \textit{vanilla learning}---and investigate their different learning dynamics. In \textit{vanilla training}, a naïve agent is trained directly with the compositional task while in \textit{primitives pre-training}, an agent first learns the primitives to expert level and then learns the compositional task (as illustrated in Figure~\ref{figure1}b).

Following the approach of~\cite{saad1995exact, biehl1995learning, patel2023rl}, we derive a set of dynamical equations that capture the learning dynamics in the high dimensional limit ($N \to \infty$). This approach allows us to describe the evolution of the high-dimensional student weights with a handful of order parameters,
\begin{equation}
\label{eq:order_params}
Q_{k} = \frac{W_k \cdot W_k}{N}, \: R_{k}=\frac{W_k \cdot W_k^*}{N}, \: S_{k}=\frac{W_k^*\cdot  W_k^*}{N}.
\end{equation}
Intuitively, $Q_{k}$ is the norm of the student primitive task $k$. Similarly, $S_{k}$ is the norm of the teacher primitive task $k$. $R_{k}$ is interpreted as the alignment or correlation between the student and teacher of the primitive task $k$. \rebuttal{The generalisation error for a single primitive task $k$ is defined by the \textit{average} disagreement between $y^*_k$ and $y_k$ and it can be described with the order parameters as follows},
\begin{equation}
\label{eq:gen_primitive_error}
\epsilon_{k,g}=\frac{1}{\pi}\cos^{-1}(R_k/\sqrt{Q_k\rebuttal{S_k}}). 
\end{equation}
Similarly, the generalisation error on the compositional task $\tilde{\epsilon}_g$ is defined by the average disagreement between $\Tilde{y}^*$ and $\Tilde{y}$ and becomes
\begin{equation}
    \label{eq: gen_error}
    \Tilde{\epsilon}_g = \frac{1}{\pi}\cos^{-1}\Biggl(\frac{\sum_{i=1}^K V^*_iV_i R_{i}}{\sqrt{\sum_{i=1}^KV_i^{*2} S_{i}} \sqrt{\sum_{i=1}^KV_i^2 Q_{i}}}\Biggr).
\end{equation}
Therefore, tracking the order parameters (Eq.~\ref{eq:order_params}) and $V_k$ allows us to characterise the generalisation error of the compositional and primitive tasks during the entire learning period as shown in Eq.~\ref{eq:gen_primitive_error}-\ref{eq: gen_error}. As we approach the high dimensional limit ($N\to\infty$), the learning curves concentrate to a typical trajectory described by the ODEs below. \rebuttal{We introduce a variable $z = \frac{n}{N}$, the ratio of the number of episodes to the input dimension, which is analogous to continuous `time' in the limit $N\to\infty$.}  for the full derivation refer to Appendix~\ref{appendix:ODE_derivation}. The equations describe the time evolution of the order parameters and $V_k$ in continuous time (z),
\begin{align}
    \label{eq: Q_ode}
    \frac{dQ_{k}}{dz} & = \frac{2\eta_w}{\sqrt{2\pi}}\Biggl(\frac{V_k^2 Q_{k}}{\sqrt{\sum_{i=1}^K{V_i^2Q_{i}}}} + \frac{V_kV^*_k R_{k}}{\sqrt{\sum_{i=1}^K{V_i^{*2}{S_{i}}}}} \Biggr )\Tilde{P}^{T-1} \nonumber \\& + \frac{\eta_w^2}{T}V_k^2 \Tilde{P}^T \\
    \label{eq: R_ode}
    \frac{dR_{k}}{dz} & = \frac{\eta_w}{\sqrt{2\pi}}\Biggl (\frac{V_k^2R_{k}}{\sqrt{\sum_{i=1}^K{V_i^2Q_{i}}}} + \frac{V_kV^*_k S_{k}}{\sqrt{\sum_{i=1}^K{V_i^{*2}S_{i}}}} \Biggr )\Tilde{P}^{T-1} \\ 
    \label{eq: V_ode}
    \frac{dV_k}{dz} & = \frac{\eta_v}{\sqrt{2\pi}}\Biggl( \frac{V_k Q_{k}}{\sqrt{\sum^K_{i=1} V_i^2 Q_{i} }} + \frac{V_k^* R_{k}}{\sqrt{\sum_{i=1}^K V^{*2}_i S_{i}}} \Biggr )\Tilde{P}^{T-1}
\end{align}

where $\Tilde{P}=1-\Tilde{\epsilon}_g$ is the probability of making a correct compositional choice. In this study, we focus on the spherical case where $\lVert V^* \rVert = \lVert V \rVert=1$ and $S_k=Q_k = 1$. This simplifies the expression of $\tilde{\epsilon}_g$ and $\epsilon_{g,k}$ and we only need to keep track of $V_k$ and $R_k$.

In the next sections we analyse evolution of the order parameters (Eqs.~\ref{eq: Q_ode}-\ref{eq: V_ode}) for \textit{primitives pre-training} and \textit{vanilla training} in detail and obtain the relevant timescales of learning.

\subsection{Primitives Pre-training}
Leveraging the compositional structure of the task, the primitives pre-training strategy aims to train a student on primitives first, and later generalise to the compositional task. During pre-training, only primitive student weights $W_{1...K}$ are updated (Eq.~\ref{eq:primitive_update}) without information from the context $V$. The compositional task is presented to the student model once it reaches expert level on the primitives $R_0$ (defined as 99\% of the maximum accuracy that the student can reach with $T$), and both $W_{1...K}$ and $V$ are updated according to Eqs.~\ref{eq:compositional_w_update}-\ref{eq:compositional_v_update}.

\underline{\textit{Timescale of primitive learning.}} Using the ODEs for a single primitive task from the RL perceptron \cite{patel2023rl}, we analytically derive a closed-form approximation of $R_k(z)$,
\begin{equation}
    \label{eq:comp_gen_rk}
    R_k(z) = -\frac{\pi}{2} + \pi\Biggl(\frac{\eta_w}{\sqrt{2\pi}\pi}(2-T)z + 2^{(T-2)} \Biggr)^{\frac{1}{2-T}}   
\end{equation}
which captures the learning timescale of a single primitive of episode length $T$ 
\begin{gather}
    \tau_{\text{primitive}} \sim \frac{1}{\eta_{w}(T-2)}2^{(T-2)}.
\end{gather}
See Appendix~\ref{appendix: asympt_primitive} for the derivation. In Figure~\ref{figure2}a, we show the learning dynamics of a single primitive from the numerical integration (solid line), approximated dynamics (dashed line) and the approximated timescale $\tau_{\text{primitive}}$ (dotted line). We can see that the approximation captures the dependence of the training timescale of a primitive task on episode length $T$. 
From this result we see that the learning timescale for a single primitive task, $\tau_\text{primitive}$, is exponential in $T$. Namely, the larger the episode length $T$, the harder the primitive task is to learn. Intuitively, this is due to sparser rewards and fewer updates in the early stages of learning which make the overall dynamics slower.

After primitives pre-training, the student is trained on the compositional task. 

\underline{\textit{Compositional generalization.}} In the compositional generalization phase of the curriculum, the student learns to adapt the acquired primitives to the compositional task. In this learning stage, both the student primitive weights $W_{1...K}$ and context $V$ are updated following Eqs.~\ref{eq:compositional_w_update}-\ref{eq:compositional_v_update}.

After primitives pre-training, each primitive reaches expert alignment $R_0$ and we can derive the approximate dynamics of $\Tilde{R}$ around $R_k \approx R_0$. This allows us to determine the dominant factor of the training time for compositional generalization,
\begin{gather}
\label{eq:tau_curriculum}
    \tau_{\text{composition}} \sim \frac{1}{\eta_v(T-1)}\tilde P_0^{2-T} \\
    \Tilde{P}_0 = \left(1-\frac{1}{\pi}\cos^{-1}\left(R_0\sum_kV_k^0V_k^*\right)\right)
\end{gather}    

where $V_k^0$ is a initial student context weight for primitive $k$. See Appendix \ref{appendix: asympt_compositional_gen} for the detailed expression and derivation.  This approximation accords well with simulation results as shown in Figure~\ref{figure2}b-c.

\underline{\textit{Altogether}}, given $K$ primitives and episode length $T$, the total training time of the \textit{primitives pre-training} curriculum 
is of the order
\begin{equation}
    \tau_{\text{curriculum}} \sim (K2^{T-2}+ \tilde P_0^{2-T}). 
\end{equation}
Since $\Tilde{P}_0 \geq 1/2$ and $K\geq2$ for the compositional task, as the primitive task complexity (episode length $T$) increases, the burden of training time mostly comes from primitive training rather than compositional generalization as clearly illustrated in Figure~\ref{figure2}d. As a consequence, after the student masters the primitives, it is relatively fast to adapt to new compositional tasks.
Note that an important role is played by the alignment of the initial context of the student $V^0$ with the target context $V^*$ as the better alignment makes larger $\Tilde{P}_0$.

\subsection{Vanilla Training}
\begin{figure*}[hbt!]
\vskip 0.2in
\begin{center}
\centerline{\includegraphics[width=\textwidth]{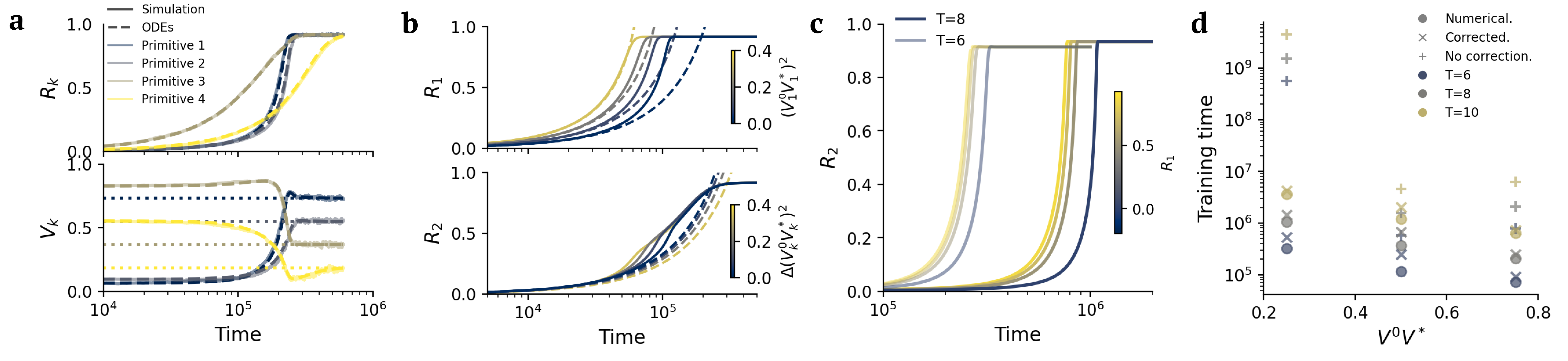}}
\caption{\textbf{Learning dynamics in vanilla learning.} \textbf{a)} Learning dynamics of each primitive task $R_k$ (top) and context $V_k$ (bottom) during vanilla training ($T=6$, $K=4$, $V^0V^*=0.5$, $\eta_w=1$, $\eta_v=1$, $N=1000$). Dynamics from simulation of the update rules Eq. \ref{eq:compositional_w_update}-\ref{eq:compositional_v_update}) (solid) are well described by the ODEs of Eq. \ref{eq: Q_ode}-\ref{eq: V_ode} (dashed). The dotted line is the target context $V^*$. \textbf{b)} $V_k^0V_k^*$ controls the delay of primitive learning. Top: Learning dynamics of the first learned primitive task ($R_1$) for varying $(V^0_1V_1^*)^2$. Smaller $V^0_1V^*_1$ leads to more delayed learning of the primitive. Bottom: Learning dynamics of the later learned primitive task $R_2$ for different $\Delta(V^0_kV_k^*)^2$ values. As the difference of $\Delta(V^0_kV_k^*)^2$ is bigger, the gap between approximation and the numerical integration is bigger. \textbf{c)} Demonstration of \textit{multiplicative learning effect} of the primitives. The second primitive learns faster as the first primitive has better knowledge. Each learning curve depicts the second primitive learning dynamics for a given initial overlap of the first primitive $R_1$ (color scale). \textbf{d)} The numerical training time (circle), approximated training time corrected for the multiplicative effect (cross), and non-corrected (plus) with respect to cosine similarity of the initial $V_0$ and the target $V^*$ and the episode length $T$. We can appreciate the importance of multiplicative effect in vanilla learning.}
\label{figure3}
\end{center}
\vskip -0.3in
\end{figure*}

In vanilla training, the student faces the compositional task immediately, updating the weights following Eqs.~\ref{eq:compositional_w_update}-\ref{eq:compositional_v_update}. 
In Figure~\ref{figure3}a, we can see that the student successfully learns the primitive tasks and context.
Notably, each primitive is learned with a different timescale in vanilla learning while the context learning for each primitive occurs with a similar timescale.

In the following section, we aim to understand the learning dynamics of primitives and context in the vanilla learning strategy. We restrict our consideration to the $K=2$ case for simplicity. 

\underline{\textit{Initial context alignment drives primitive learning speed.}}
In vanilla learning, the learning dynamics is much more convoluted due to the coupling of the order parameters $R$ and $V$. Yet, we can still capture important factors which determine the early learning timescale for each primitive in vanilla learning. In this section, we first show how the alignment of the initial context $V^0$ and the target context $V^*$ modulates the learning time of the primitives, explaining the different learning timescale of each primitive observed in Figure~\ref{figure3}a.

In the early phase of vanilla learning, we can assume that $R_{k'\neq k} \approx 0, \Tilde{R} \approx R_kV_kV_k^*\ll1$. Under these assumptions, we approximate the early dynamics of $R_k$ as follows:
\begin{equation}
    \frac{dR_k}{dz} \approx \frac{\eta_w}{\sqrt{2\pi}} V_k^0V_k^*\Bigl(\frac{1}{2}+\frac{V_k^0V_k^*R_k}{\pi}\Bigr)^{T-1}.
    \label{eq:vanilla_approx_1stk}
\end{equation}
Solving the ODE above, we get a closed-form approximation of early $R_k$ dynamics and the characteristic time in vanilla learning as following, 
\begin{align}
    R_k(z) = \frac{1}{V_k^0V_k^*}\Bigl(-\frac{\pi}{2}+&\pi \Bigl(\frac{\eta_w}{\sqrt{2\pi}\pi}(2-T)\Bigl(V_k^0V_k^*\Bigr)^2z \nonumber \\&+ 2^{T-2}\Bigr)^{\frac{1}{2-T}} \Bigr)
    \label{eq:vanilla_R}
\end{align}
\begin{equation}
\label{eq:z_vanilla1}
    \tau_{\text{vanilla},k} \sim \frac{1}{\eta_w(T-2)}2^{(T-2)}\Bigl(\frac{1}{V_k^0V_k^*}\Bigr)^2.
\end{equation}

From Eq.~\ref{eq:z_vanilla1} we can observe that the learning timescale of primitive $k$ is attenuated by a $1/(V_k^0 V_k^*)^2$ term compared to the primitive training shown in Eq.~\ref{eq:comp_gen_rk}. In other words, each primitive has a different timescale for learning determined by $1/(V_k^0 V_k^*)^2$.

Intuitively, $V_k^0$ represents the initial attention of the model to the primitive $k$. $V_k^*$ represents the actual relevance of the primitive $k$ in the compositional task. 
Hence higher $V^0_kV^*_k$ means that the primitive stands out during learning and is in fact important, leading to faster learning. Given that the context vector has norm 1, the student is forced to trade-off in which direction to put more relevance or attention. 
For sake of comprehension, we denote the first trained primitive as $k=1$ and the later trained one as $k=2$ in the following. In Figure~\ref{figure3}b, we can clearly see that the approximation holds for $V_kV_k^*$ sufficiently apart. Nevertheless it fails when the difference between $\Delta (V^0V^*)^2 = (V^0_1V^*_1)^2 - (V^0_2V^*_2)^2$ becomes small. In the next section, we further investigate the cause of this gap, which brings us to a \textit{multiplicative effect} in primitive learning during vanilla learning.

\underline{\textit{Emergence of multiplicative primitive learning.}}
So far in considering $R_k$, we assume $R_{k'}\approx0, k \neq k'$. However, as learning progresses, the overall $\Tilde{R}$ becomes larger due to the contribution of $R_{k'}$, and this accelerates the learning of primitive $k$. This effect explains why the approximation for $R_{k=1}$ predicts slower learning than the actual dynamics, and why the gap is bigger when $\Delta V^0V^*$ is smaller (Figure~\ref{figure3}b top). On the other hand, from the perspective of $R_2$, large $\Delta (V^0V^*)^2$ indicates that $R_1$ already has been significantly increased, and causes a strong divergence from the numerical integration (Figure~\ref{figure3}b bottom).

These observations indicate that learning one primitive accelerates learning of the rest of the primitives. Interestingly, despite the task design where all primitives are modularised and independent, the learning of one primitive increases $\Tilde{R}$ and leads to rapid learning of the other primitives. We probe this multiplicative primitive learning effect in vanilla compositional task learning with a controlled experiment. We initialise the first primitive $R_1$ at different levels and simulate the learning dynamics of the second primitive task under the vanilla learning strategy. As shown in Figure.~\ref{figure3}c, a high initial value of $R_1$ leads to fast learning of the second primitive task, $R_2$.   

In the next section, we derive the timescale for vanilla learning based on our understanding of attenuation from $V_0V^*$ and the multiplicative effect.

\underline{\textit{Timescale of vanilla learning.}}
To derive the approximate total learning timescale of vanilla learning, we consider the primitive which is learned last ($R_2$), since the last task to be learnt will decide the total compositional task training timescale. We take the multiplicative effect caused by the first primitive into account and add a correction term from the contribution of $R_1$ in Eq.~\ref{eq:vanilla_R}. See Appendix~\ref{appendix: asympt_vanilla} for the full derivation. In essence, the total learning time of vanilla learning can be described as
\begin{equation}
\tau_{\text{vanilla}}^{(K=2)} \sim \frac{1}{\eta_w(T-2)}2^{(T-2)}\frac{1}{(V^0_1V_1^*)^2+(V^0_2V_2^*)^2}. 
\label{eq:tau_vanilla}
\end{equation}
The correction from the multiplicative effect is critical in determining the total learning time scale and accounting for it provides good agreement with the total training time in vanilla learning obtained from numerical integration as shown in Figure.~\ref{figure3}d.

\section{Curriculum Learning Effects}
\subsection{Improved Learning Speed}
\label{subsection:learning_speed}
\begin{figure}[ht]
\vskip 0.2in
\begin{center}
\centerline{\includegraphics[width=\columnwidth]{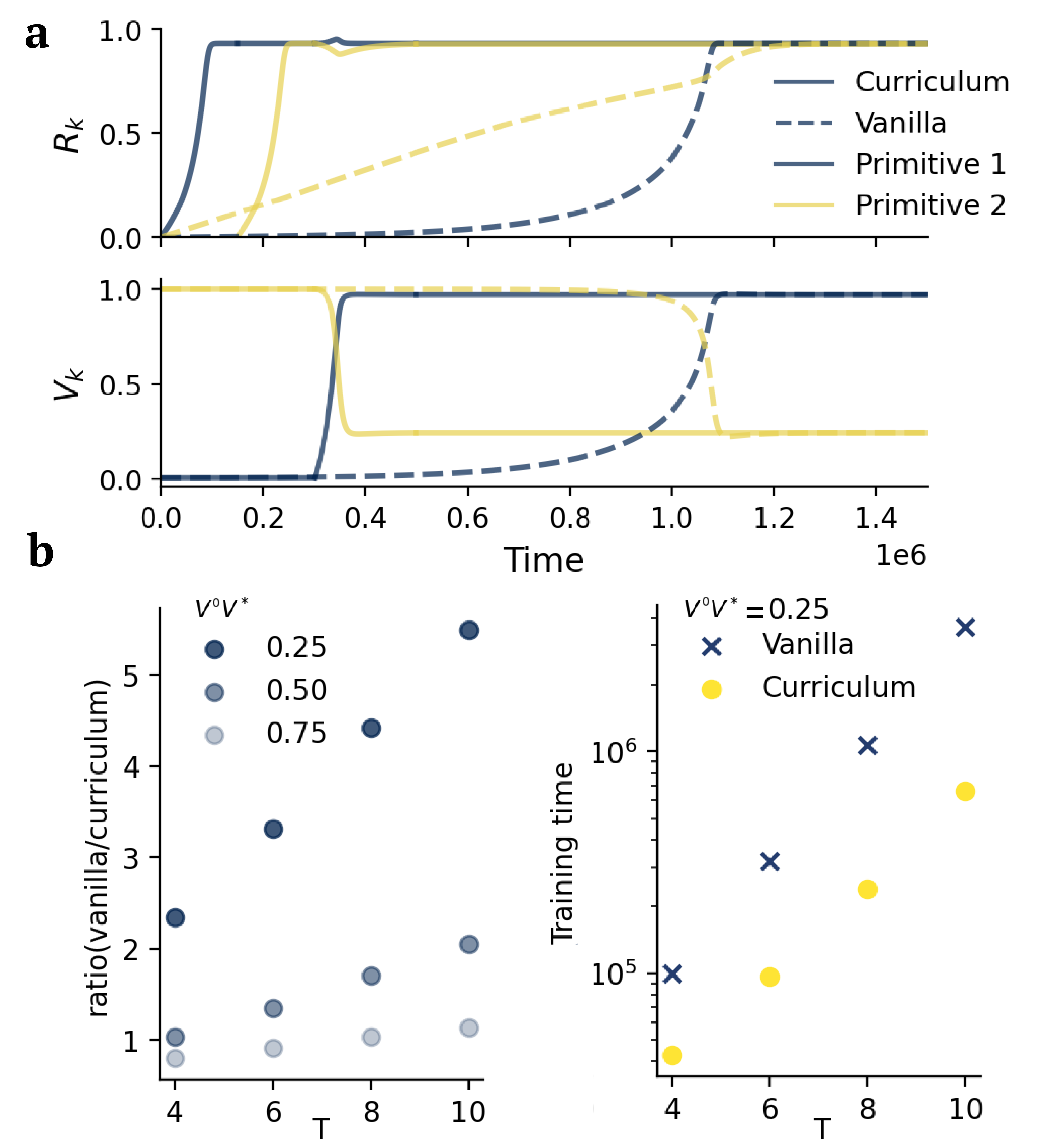}}
\caption{\textbf{Speed boost from curriculum learning.} \textbf{a)} Example of faster learning under the primitives pre-training strategy (solid) compared to vanilla learning (dashed) ($K=2$, $V^0V^*=0.25$, $\eta=1$, $N=1000$). \textbf{b)} Left) The ratio between training time of vanilla learning and primitive pre-training curriculum learning from numerical simulation. Different colors indicate the initial cosine similarity of the context weights $V^0V^*$ ($K=2$). Right) The numerical learning time from different strategies ($V^0V^*=0.25$).}
\label{figure4}
\end{center}
\vskip -0.5in
\end{figure}

Finally, after analysing the learning time for the two different curricula strategies, we compare the learning time across $T$ and the alignment between the initial and the target context $V^0V^*$ for $K=2$. We find that primitive pretraining can offer substantial learning speed benefits compared to vanilla training. Figure~\ref{figure4}a shows an example where primitive pre-training (solid line) is much faster than vanilla training (dashed line). 
Turning in more detail to the effect of primitive difficulty, we can see that increasing $T$ leads to exponential growth of training time in both vanilla learning and primitive pre-training (Figure~\ref{figure4}b right), while their growth rate differs.
Furthermore, as we can infer from Eq.~\ref{eq:tau_curriculum} and Eq.~\ref{eq:tau_vanilla}, having larger $T$ (higher task complexity) and smaller cosine similarity between $V^0$ and $V^*$ significantly increases the learning speed boost from the primitives pre-training curriculum (Figure~\ref{figure4}b left). See Appendix~\ref{appendix:additional} and Figure~\ref{figureA1} for a similar tendency in $K=4$ case and under an extended learning rate hyperparameter search. Hence primitives pretraining confers a learning speed advantage in our compositional task setting. 

\subsection{Improved Noise Robustness}
\label{subsection:learning_speed}
In this section we investigate another potential benefit of primitives pre-training curriculum: robustness to the noise in training. In the real world, learning is not noise-free. In the presence of noise during learning, we study the robustness of curriculum learning in comparison to vanilla learning. 

\begin{figure}[ht]
\vskip 0.2in
\begin{center}
\centerline{\includegraphics[width=\columnwidth]{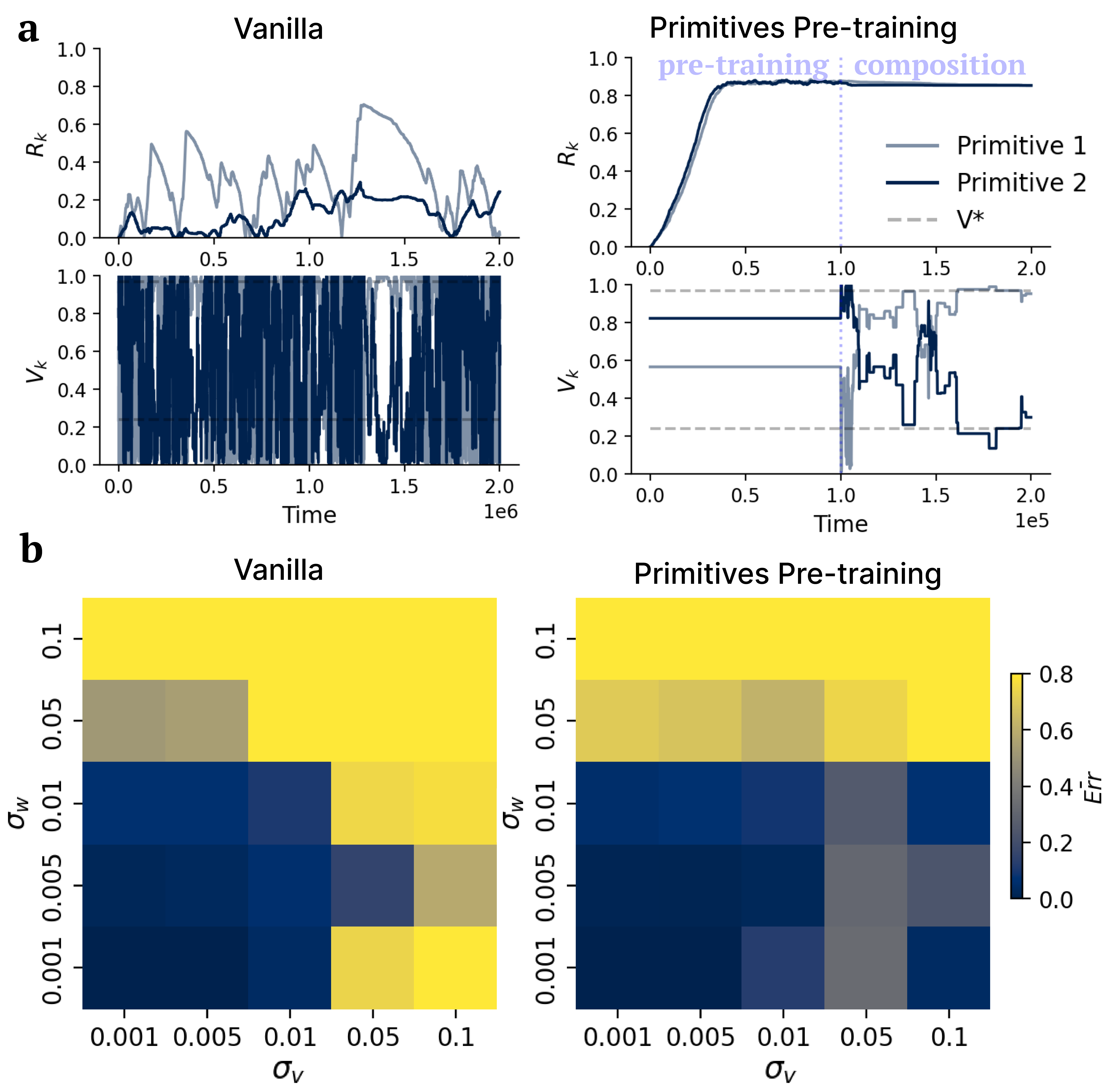}}
\caption{\textbf{Robustness to noise in curriculum learning.} \textbf{a)} Example of $R_k$ and $V_k$ dynamics during noisy learning ($K=2$, $V^0V^*=0.75$, $\sigma_w=0.01$, $\sigma_v=0.1$), in vanilla training and primitives pre-training. \textbf{b)} Average difference of $\Tilde{R}_0 - \Tilde{R}_\text{noise}$ during the last 5000 iterations of training in two different training protocols across a range of $\sigma_w$ and $\sigma_v$($K=2$, $V^0V^*=0.25$). Primitives pre-training is more robust compared to vanilla training especially when noise in the context update $\sigma_v$ is large.}
\label{figure5}
\end{center}
\vskip -0.4in
\end{figure}

We inject i.i.d. Gaussian noise $\varepsilon_w\sim\mathcal{N}(0,\sigma_w)$ and $\varepsilon_v\sim\mathcal{N}(0,\sigma_v)$ into each element of the gradient of $W$ and $V$, respectively, and compare the learning efficiency in simulation of the two training protocols. We varied the noise levels $\sigma_{w}$ and $\sigma_{v}$, and compare the average deviation of $\Tilde{R}_\text{noise}$ from the $\Tilde{R}_0$ which is reachable without noise during the training. We allow both training protocols enough training iterations to reach $\Tilde{R}_0$ in the noise-free case. We add the noise term to the updates throughout training. 
Figure~\ref{figure5}b shows the difference in learning performance of the training protocols as we vary the noise scales $\sigma_w$ and $\sigma_v$. 
When $\sigma_w$ is large, both vanilla learning and curriculum learning suffer and fail to learn the compositional task ($\sigma_w\geq0.05$ in the heatmap) irrespective of $\sigma_v$. When both $\sigma_w$ and $\sigma_v$ are small ($\sigma_w<0.05, \sigma_v<0.05$ in the heatmap), both learning protocols successfully learn the task. An interesting range is where $\sigma_w$ is small but  $\sigma_v$ is relatively large ($\sigma_w\leq0.05, \sigma_v\geq0.05$ in the heatmap). In this range, primitives pre-training provides significantly better learning than vanilla training. In Figure~\ref{figure5}a ($\sigma_w=0.01$, $\sigma_v=0.1$), we illustrate an example case where primitives pre-training curriculum enables learning of all primitives and the composition context while vanilla learning fails. 

The robustness to noisy training in structured compositional tasks under curriculum learning comes from separating the two sources of noise: primitives learning and compositional generalization. In the primitives pre-training curriculum, the student initially only needs to handle the noise in the primitives. After learning the primitives, the effective noise only comes from the context learning, since the primitives are already well-aligned to the teacher and the fluctuations for the primitives are small. In contrast, in vanilla learning, the student needs to learn the primitive task and context learning simultaneously with mixed noise sources which makes learning more challenging.

\section{Why do animals need shaping?}
In psychology and neuroscience, there has been a longstanding appreciation of the importance of \textit{behaviour shaping} for animal learning~\cite{pavlov1927conditioned,skinner2019behavior}. In behaviour shaping, an animal learns multiple components of a complex task incrementally. It is colloquially known that shaping is critical for animals to learn complex tasks: different shaping strategies lead to different training times and sometimes animals completely fail to learn a task without shaping.

In our model, the primitives pre-training curriculum is reminiscent of shaping procedures. During the primitives pre-training phase, the student learns sub-components independently before combining them in the right context to solve the compositional task. Our analysis of the learning dynamics shows that the learning time for curriculum learning is significantly shorter than vanilla learning when the primitive task complexity ($T$) is high and the initial and target composition context alignment $V^0V^*$ is poor. This observation addresses several interesting findings from psychology and neuroscience. Many works in human psychology report feature-level biases on which features of objects or tasks draw attention and thus are easier to learn~\cite{landau1988importance, diesendruck2003specific,hudson2005regularizing}. In rodent studies, similar notions are known as attentional bias or rule bias~\cite{smith2012implicit,vermaercke2014more,goltstein2021mouse}. These studies indicate that finding the right combination of importance or attention (context) poses a great challenge for learning the composition of multiple tasks, especially when it is largely misaligned to the initial bias or the search space is large. In this context, our analysis stresses that shaping or curriculum learning helps acquire the relevant primitives independently from the initial bias or attention toward certain features. Studies in humans also found that primitives pre-training can dramatically affect generalisation, allowing subjects to generalise after few trials under the right curriculum protocol~\cite{dekker2022curriculum,eckstein2021mind}. Furthermore, our finding and analysis of the \textit{multiplicative effect} implies that prior knowledge is important to further learning dynamics and behaviour. This aspect of learning in humans and animals has been observed in many works~\cite{vergnaud1994multiplicative,block1982assimilation,akrami2018posterior}. Although in our curriculum learning setting the training order among the primitives does not matter because all primitives are independent and orthogonal, in more realistic settings where the sub-tasks are correlated, our results suggest that the order of pre-training will impact the overall result of the shaping procedure. 
\section{Discussion}
In this study, we provide a theory of a simple case of task composition and curriculum learning. By formulating a compositional task with primitives and compositional context in the teacher-student setup, we derive a set of ODEs that describe the learning dynamics of the task. This allows us to analytically study the distinct learning dynamics emerging in two different curricula, namely primitives pre-training and vanilla training. In our setting, we characterise potential benefits of curriculum learning: a speed boost in learning, and robustness to the noise during learning. Our model provides a quantitative understanding of the importance of shaping in learning compositional tasks.  

\rebuttal{While the scope of this study is limited to a particular type of compositionality, namely the linear combination of sub-tasks, many complex tasks can be described with different kinds of compositionality including sequential chaining of sub-tasks or recursive composition in tree-like task structures. The analytical study of learning dynamics in different kinds of compositional structure is an important avenue for future work.} Our study further assumes strict modularisation of the primitives. By relaxing this constraint, we might observe more diverse behaviours of the model. One interesting future direction is to introduce a notion of task similarity either in terms of the primitives rules~\cite{lee2021continual, pmlr-v162-lee22g} or their input data~\cite{mannelli2022unfair}. In our current model, the primitives are independently modularised and orthogonal to each other. We speculate that the benefits of curriculum learning that our current model shows is a lower bound on the actual potential effect. For example, correlation between the primitives might ease pre-training since one primitive might be able to leverage already acquired knowledge from a correlated primitive. Furthermore, non-modularized primitives and a compositional task built on them can possibly find alternative solutions during vanilla learning, which might harm compositional generalization performance on a series of different composition contexts. A broader study on different configurations of compositional tasks and primitives is an exciting future direction. 

\section*{Acknowledgement}
We thank Athena Akrami, Ann Chunyu Duan, Maria Eckstein, Kishore Kuchibhotla and Nishil Patel for useful discussions. SSM acknowledges Bocconi University for hospitality during the final stage of this project. 
This work was supported by a Sir Henry Dale Fellowship from the Wellcome Trust and Royal Society (216386/Z/19/Z) to AS, and the Sainsbury Wellcome Centre Core Grant from Wellcome (219627/Z/19/Z) and the Gatsby Charitable Foundation (GAT3755). 

\section*{Impact Statement}
This paper focuses on theoretical research, and currently,
we do not envision direct societal impact emerging from its
conclusions.

\bibliography{example_paper}
\bibliographystyle{icml2024}

\newpage
\appendix
\onecolumn

\section{Summary of RL perceptron and Derivation of the Learning Rules }
\label{appendix:rl_perceptron}
\rebuttal{The RL perceptron~\cite{patel2023rl} is a particular instantiation of the teacher-student framework designed to capture the learning of RL agents. In this case, the labels provided by the teacher are interpreted as the correct choice that the student must make to receive a reward. More precisely, given an input $x_{t}$ (randomly sampled vector) at time step $t$, the teacher network $W^*$ will generate the correct binary choice $y_t^* = \textrm{sign}(W^*\cdot x_t/\sqrt{N})$. The student network $W$ will receive the same input $x_t$ and make its prediction $y_t = \textrm{sign}(W\cdot x_t/\sqrt{N})$. If all $T$ decisions in the episode are correct, the student will receive a reward and update its weights $W$ according to the REINFORCE policy gradient update rule~\cite{sutton1999policy}. The equation for the update is given by Eq.~\ref{eq:primitive_update}. Patel et al.(\citeyear{patel2023rl}) showed that this perceptron-like learning rule is like policy gradient learning and it is amenable to analytical characterisation.}

\rebuttal{
In more detail, consider a probabilistic policy function $\pi_{W}$ such that $\pi_{W}$ samples a binary decision ($y_t = \pm1$) based on input $x_t$ at each time step $t (t=1…T)$ following the policy $ \pi_{W}(y_t|x_t) = \frac{1}{1+\exp{(-y_t(W\cdot x_t)/\sqrt{N})}}$, where $N$ is input dimension. The reward is only available when decisions over all time steps ($t=1…T$) are correctly given so the expected reward becomes $J = \sum^T \log(\pi_W(y_t|x_t))$. Following the update rule of \citet{patel2023rl}, we take the gradient of the expected reward $J$ with respect to policy parameter $W$, and with large input dimension $N$, we get the learning rule Eq.~\ref{eq:primitive_update}.}

\rebuttal{
Similarly, Eq.~\ref{eq:compositional_w_update}-\ref{eq:compositional_v_update} are results of policy gradient learning on compositional policy. Instead of a single policy for one primitive, the compositional policy becomes $\pi_{W,V_{1…K}}(y_t|x_t)=\frac{1}{1+ \exp(-y_t\sum_K{V_kW_k\cdot x_k/\sqrt{N}})} $ with $K$ primitives. Replace this policy for the expected reward $J = \sum^T \log(\pi_W(y_t|x_t))$ and taking gradient with respect to the policy parameters $W_k$ and $V_k$ leads to the  update rules in Eq.~\ref{eq:compositional_w_update}-\ref{eq:compositional_v_update}.}

\section{Derivations of ODEs}
\label{appendix:ODE_derivation}

In the thermodynamic limit, the stochastic updates of the student's dynamics concentrate to a deterministic limit, leading to effective dynamics characterised by ODEs. In this section, we show how to derive Eqs.~\ref{eq: Q_ode}-\ref{eq: V_ode} described in the main text.

We define the alignment fields of teacher and student weights of each primitive task $k$ as $\nu_{k} = \frac{W_k^* \cdot x_k}{\sqrt{N}}$ and $\lambda_{k} = \frac{W_k \cdot x_k}{\sqrt{N}}$, and the compositional alignment as follows,
\begin{gather}
\Tilde{\nu}= \sum_{i=1}^K{V_i^*\frac{W^*_i \cdot x_{i}}{\sqrt{N}}} = \sum_{i=1}^K{V_i^* \nu_{i}}\\
\Tilde{\lambda}= \sum_{i=1}^K{V_i\frac{W_i \cdot x_{i}}{\sqrt{N}}} = \sum_{i=1}^K{V_i \lambda_{i}}
\end{gather}
notice that the primitive decision and compositional task decision can be expressed with the alignment fields: $y_k^*=\textrm{sign}(\nu_k), y_k=\textrm{sign}(\lambda_k), \Tilde{y}^*=\textrm{sign}(\Tilde{\nu})$ and $\Tilde{y}=\textrm{sign}(\Tilde{\lambda})$. Note that these quantities are time-dependent and we will add the subscript $t$ to indicate the time step when necessary.

These fields are key stochastic variables of the problem, indeed the fundamental idea behind the ODE limit is to track their distribution at all times. It is easy to notice that they are centered Gaussian variables with covariance given by the order parameters (Eq.~\ref{eq:order_params}) defined in the paper.  

Redefining the expectations in terms of these fields allows us to derive fundamental observables of the problems, for instance the average probability of success $\Tilde{P}$ (i.e. $\Tilde{y_t} = \Tilde{y^*_t}$). This quantity is equivalent to the probability of having both, the ground truth and the student choices, lying on the same hypersector defined as $\Tilde{\nu}$ and $\Tilde{\lambda}$,
\begin{gather}
\Tilde{P} = \Big\langle1 - \frac{\cos^{-1}({\Tilde{\nu} \Tilde{\lambda}})}{\pi}\Big\rangle.
\end{gather}

Given these preliminary definitions, the goal is to derive the updates for $Q_k$, $R_k$, and $V_k$. 

\paragraph{Useful equalities.} The following analysis will need to evaluate some Gaussian averages. It is therefore convenient to compute some useful relations:
\begin{align}
    \label{eq:averages1}
    & \langle\nu_k sign(\Tilde{\nu})\rangle = \sqrt{\frac{2}{\pi}}{\frac{{V_k^* S_{k}}}{\sqrt{\sum^K_k V_k^{*2} S_{k}}}},
    \\
    & \langle\lambda_k sign(\Tilde{\nu})\rangle = \sqrt{\frac{2}{\pi}}\frac{V_k^* R_{k}}{\sqrt{\sum_k^KV_k^{*2} S_{k}}},
    \label{eq:averages2}
    \\
    & \langle\lambda_k sign(\Tilde{\lambda})\rangle = \sqrt{\frac{2}{\pi}}{\frac{V_k Q_{k}}{\sqrt{\sum_k^K{V_k^2 Q_{k}}}}}.
    \label{eq:averages3}
\end{align}

Notice that the correct decision, $sign\tilde\nu=sign\tilde\lambda$, is equivalent to $\theta(\tilde\nu\tilde\lambda)$ with $\theta$ the Heaviside function.
It is useful to notice that following relations hold
\begin{align}
sign(\Tilde{\lambda})\theta(\Tilde{\nu}\Tilde{\lambda}) = sign(\Tilde{\nu})\theta(\Tilde{\nu}\Tilde{\lambda}) = \frac{1}{2}(sign(\Tilde{\lambda}) + sign(\Tilde{\nu})).
\end{align}
The average of these equalities can be evaluated in different ways, an easy approach is to evaluate the two fields independently by conditioning and using the properties of Guassian variables. 

\paragraph{ODEs.} 
Starting from the definition of the gradient updates Eqs.~\ref{eq:primitive_update}-\ref{eq:compositional_v_update}, it has been shown that the updates for the order parameters concentrate to their typical value as $N\to\infty$~\cite{goldt2019dynamics}. Therefore,
to derive dynamics of $V_k$, we evaluate the average Eq.~\ref{eq:compositional_v_update} over the random fields, we use the angular brackets to denote the average over all the random variables. 
\begin{align}
    \frac{dV_k}{dz} &= \eta_v\Biggl \langle \frac{1}{T} \sum^{K,T}_{i,t=1}\Tilde{y_t}\frac{W_i \cdot x_{i,t}}{\sqrt{N}}\textit{I}(\Phi)\Biggr \rangle \nonumber  \\& = \eta_v\Biggl\langle\lambda_k sign(\Tilde{\lambda}) \theta(\Tilde{\nu}\Tilde{\lambda}) \Biggr \rangle \Tilde{P}^{T-1} \\& = \frac{\eta_v}{\sqrt{2\pi}}\Biggl( \frac{V_k Q_{k}}{\sqrt{\sum^K_{i=1} V_i^2 Q_{i} }} + \frac{V_k^* R_{k}}{\sqrt{\sum_{i=1}^K V^{*2}_i S_{i}}} \Biggr )\Tilde{P}^{T-1}.
\end{align}

From the definition of $R_k$ in Eq.\ref{eq:order_params}, we take the dot product by $W_k^*/N$ on both sides of Eq. \ref{eq:compositional_w_update} and average at the $m\text{th}$ update.
\begin{align} 
    {R_k^{m+1}-R_k^{m}} = \frac{\eta_w}{N} \Biggl \langle \frac{1}{T} \sum^T_{t=1}{V_k sign(\Tilde{\lambda})\frac{ W_k^* \cdot x_{t,k}}{\sqrt{N}}} \Biggr \rangle
\end{align}
finally in the thermodynamic limit
\begin{align}
    \frac{dR_k}{dz} & = \eta_w \Biggl \langle \frac{1}{T} \sum^T_{t=1}{V_k sign(\Tilde{\lambda})\nu_k \textit{I}(\Phi)} \Biggr \rangle \nonumber  = \frac{\eta_w}{2}V_k\Bigl \langle \nu_k sign(\Tilde{\lambda}) + \nu_k sign(\Tilde{\nu}) \Bigr \rangle \Tilde{P}^{T-1} \\& =\frac{\eta_w}{\sqrt{2\pi}}\Biggl (\frac{V_k^2R_{k}}{\sqrt{\sum_{i=1}^K{V_i^2Q_{i}}}} + \frac{V_kV^*_k S_{k}}{\sqrt{\sum_{i=1}^K{V_i^{*2}S_{i}}}} \Biggr )\Tilde{P}^{T-1}.
    \label{eq: R_ode}
\end{align}

Since $Q_k={W_k}\cdot{ W_k}/N$, we take the square of Eq.~\ref{eq:compositional_w_update} and, after taking the thermodynamic limit, we obtain
\begin{align}
    \frac{dQ_k}{dz} &= 2\eta_w\Biggl \langle \frac{1}{T} \sum^T_{t=1}V_k sign(\Tilde{\lambda})\lambda_k\Biggr \rangle \Tilde{P}^{T-1} \nonumber + \frac{\eta_w^2}{T^2}\Bigl \langle \sum^{T,T}_{t,t'}{V_k V_k \Tilde{y_t}\Tilde{y_{t'}}x_t^\mathsf{T} x_{t'}} \Bigr \rangle \Tilde{P}^{T}.
    \\
    &= 2\eta_w\Biggl \langle \frac{1}{T} \sum^T_{t=1}V_k sign(\Tilde{\lambda})\lambda_k\Biggr \rangle \Tilde{P}^{T-1} \nonumber + \frac{\eta_w^2}{T} V_k^2 \Tilde{P}^{T} \\&=\frac{2\eta_w}{\sqrt{2\pi}}\Biggl(\frac{V_k^2 Q_{k}}{\sqrt{\sum_{i=1}^K{V_i^2Q_{i}}}} + \frac{V_kV^*_k R_{k}}{\sqrt{\sum_{i=1}^K{V_i^{*2}{S_{i}}}}} \Biggr )\Tilde{P}^{T-1} \nonumber + \frac{\eta_w^2}{T}V_k^2 \Tilde{P}^T.
\end{align}
where the last steps follows from the fact the correlation between $x$ and the fields becomes negligible to the leading order and we can average them following the Gaussianity.

Finally, for the \textit{spherical case} (ie. $||W_k||=1\;\forall k$), we obtain the following expression that can be derived by using a Lagrange multiplier to enforce the spherical constraint
\begin{align}
\frac{dR_k}{dz}=\frac{\eta_w}{\sqrt{2\pi}}V_kV_k^*(1-R_k^2)(1-\cos^{-1}(\tilde{R}))^{T-1} + \frac{\eta_w^2}{T}V_k^2(1-\cos^{-1}(\tilde{R}))^T.
\label{eq:sphericalR}
\end{align}

\section{Asymptotic Analysis of Learning Dynamics}
\subsection{Primitives learning}
\label{appendix: asympt_primitive}
From the ODE of a single RL perceptron \cite{patel2023rl}
\begin{align}
\frac{dR_k}{dz} &= \frac{\eta_w}{\sqrt{2\pi}}(1-R_k^2)\Biggl( 1-\frac{1}{\pi}\cos^{-1}(R_k)\Biggr)^{T-1}-\frac{\eta_w^2}{2T}R_k\Biggl(1-\frac{1}{\pi}\cos^{-1}(R_k)\Biggr)^T,
\end{align}
we approximate the function around the point $R_k \approx 0$ to capture the early dynamics and get
\begin{align}
    \frac{dR_k}{dz} & \approx \frac{\eta_w}{\sqrt{2\pi}}\Biggl(1-\frac{1}{\pi}\cos^{-1}(R_k)\Biggr)^{T-1} \nonumber \approx \frac{\eta_w}{\sqrt{2\pi}}\Biggl(\frac{1}{2} + \frac{R_k}{\pi}\Biggr)^{T-1}.
\end{align}

Solving the ODE above, we get a closed form expression on $R_k$ dynamics as follows,
\begin{align}
R_k = -\frac{\pi}{2} + \pi\Biggl(\frac{\eta_w}{\sqrt{2\pi}\pi}(2-T)z + 2^{(T-2)} \Biggr)^{\frac{1}{2-T}}.
\end{align}

\subsection{Compositional Generalisation after pre-training }
\label{appendix: asympt_compositional_gen}
\rebuttal{Since we consider a case where $\lvert V \rvert =1$, we use a Lagrange multiplier to impose the constraint on our $V_k$ update
\begin{align}
\label{eq:lagrangian_V}
   \frac{dV_k}{dz} & = g_k - \lambda V_k \\
   \label{eq:multipler}
   \sum_k V_kg_k - \lambda & = 0 \Rightarrow \lambda = \sum_k V_k g_k.
\end{align}
}

\rebuttal{Plugging Eq.~\ref{eq:multipler} into Eq.~\ref{eq:lagrangian_V} and rearranging the equation for $g_k$:
\begin{align}
    \label{eq:constratint}
    g_k=\frac{\eta_v}{\sqrt{2\pi}}\Bigl(V_k + V_k^*R_k\Bigr)\Bigl(1-\frac{1}{\pi}\cos^{-1}(\Tilde{R})\Bigr)^{T-1}.
\end{align}
We obtain $V_k$ update with the constraint satisfied by plugging in Eq.~\ref{eq:constratint} into Eq.~\ref{eq:lagrangian_V} and $\Tilde{R}$, and by multiplying $R_kV_k^*$ on both sides:
\begin{gather}
    \frac{dV_k}{dz} = \frac{\eta_v}{\sqrt{2\pi}}\Bigl(V_k + V_k^*R_k\Bigr)\Bigl(1-\frac{1}{\pi}\cos^{-1}(\Tilde{R})\Bigr)^{T-1}-\frac{\eta_v}{\sqrt{2\pi}}V_k\Bigl(1 + \Tilde{R}\Bigr)\Bigl(1-\frac{1}{\pi}\cos^{-1}(\Tilde{R})\Bigr)^{T-1}\\
    \label{eq:compositionalR_constraint}
    \frac{d\Tilde{R}}{dz} = \frac{\eta_v}{\sqrt{2\pi}}\Bigl(1-\frac{1}{\pi}\cos^{-1}(\Tilde{R})\Bigr)^{T-1}\Bigl(\sum_k V_k^2R_k^2 - \Tilde{R}^2\Bigr).
\end{gather}
}

\rebuttal{After the pre-training of all primitives, we can assume $R_k \approx R_0$ to be a constant, where $R_0$ is a maximum $R$ which can be achieved with the episode length $T$. This allows us to have the initial value of $\Tilde{R_0}\approx\sum_kV^*_kV_kR_0$ at the beginning of compositional generalisation $z_{compositional} = 0$. We linearize Eq.~\ref{eq:compositionalR_constraint} around $\Tilde{R} = \Tilde{R}_0$ treating $V_k$ as constant and get
\begin{align}
    \label{eq:compositional_approx_ode}
    \frac{d\Tilde{R}}{dz}=  \frac{\eta_v}{\sqrt{2\pi}}\Tilde{P}_0^{T-1}\Bigl(1+\frac{\tilde{R}_0(T-1)}{\Tilde{P}_0\pi\sqrt{1-\Tilde{R}_0^2}}\Bigr)\Bigl(R_0^2 + \Tilde{R_0}^2-2\Tilde{R}_0\Tilde{R}\Bigr)\Bigl(1-(\frac{\Tilde{P_0}\pi\sqrt{1-\Tilde{R}_0^2}}{(T-1)}+\Tilde{R}_0)^{-1}\Tilde{R}\Bigr).
\end{align}}

\rebuttal{For convenience, we define the below variables
\begin{align}
\alpha = & \frac{(T-1)}{\Tilde{P_0}\pi\sqrt{1-\Tilde{R}_0^2}} \\
a= &-\frac{2\Tilde{R}_0\eta_v}{\sqrt{2\pi}}\Tilde{P}_0^{T-1}(1-\alpha\Tilde{R}_0)\\
b=& -\frac{1}{2\tilde{R}_0}\bigl(R_0^2 + \Tilde{R}_0^2\bigr) \\
c= & \alpha(1-\alpha\Tilde{R}_0)^{-1}.
\end{align}
With the above variables, we rearrange Eq.~\ref{eq:compositional_approx_ode}
\begin{align}
\frac{d\Tilde{R}}{dz} = a(b+\Tilde{R})(1+c\Tilde{R}).
\end{align}
}

\rebuttal{Solving this, we get approximated closed form solution of $\Tilde{R}(z')$ after primitive pre-training where $z'=z-z_{\text{pre-training}}$,
\begin{align}
\Tilde{R}(z')=\frac{Ae^{a(1-bc)z'}-b}{1-Ace^{a(1-bc)z'}},\quad A = \frac{\Tilde{R}_0+b}{1+c\Tilde{R}_0}.
\end{align}}

\rebuttal{Since we are looking for a time $z'$ where $V$ and $V^*$ gets aligned $(V \cdot V^*=1)$ and $\Tilde{R}$ recovers possible maximum value,, we can solve $\Tilde{R}(z')=R_0$ and get 
\begin{gather}
z'_{(\Tilde{R}=R_0)} = \frac{\sqrt{2\pi}}{\eta_v}\Tilde{P}_0^{1-T}\Bigl(\alpha(\Tilde{R}_0^2-R_0^2)-2\Tilde{R}_0\Bigr)^{-1}\log(\frac{R_0+b}{A(cR_0+1)}).
\end{gather}
We can characterise the critical timescale of compositional learning phase after primitives pre-training $z'_{(\Tilde{R}=R_0)}$ with the dominant factor $\Tilde{P}_0^{2-T}$ with respect to the task difficulty $T$.}

\subsection{Vanilla Learning Time}
\label{appendix: asympt_vanilla}
We approximate the dynamics of $R_k$ by considering $R_{k'\neq k}$ components to account for the multiplicative learning effect. We modify Eq.~\ref{eq:vanilla_approx_1stk} by adding the contribution of the other primitives $k\neq k'$ as follows for $R_1$ and $R_2$

\begin{gather}
    \frac{dR_1}{dz} \approx \frac{\eta_w}{\sqrt{2\pi}} V_1^0V_1^*\Bigl(\frac{1}{2}+\frac{V_1^0V_1^*R_1 + V_2^0V_2^*R_2} {\pi}\Bigr)^{T-1}\\
    \frac{dR_2}{dz} \approx \frac{\eta_w}{\sqrt{2\pi}} V_2^0V_2^*\Bigl(\frac{1}{2}+\frac{V_2^0V_2^*R_2 + V_1^0V_1^*R_1} {\pi}\Bigr)^{T-1}.
    \label{eq:vanillaRz_approx2_correction}
\end{gather}

Finally, we get
\begin{gather}
    \label{eq:R1_correction_sol}
    R_1(z) = \frac{1}{A}\Bigl(-\frac{\pi}{2}+\pi \Bigl(\frac{\eta_w}{\sqrt{2\pi}\pi}(2-T)(V_1^0V_1^*)Az + 2^{T-2}\Bigr)^{\frac{1}{2-T}} \Bigr), \\
    \label{eq:R2_correction_sol}
    R_2(z) = \frac{1}{B}\Bigl(-\frac{\pi}{2}+\pi \Bigl(\frac{\eta_w}{\sqrt{2\pi}\pi}(2-T)(V_2^0V_2^*)Bz + 2^{T-2}\Bigr)^{\frac{1}{2-T}} \Bigr),
\end{gather}
with $A = V^0_1V_1^* + CV_2^0V_2^*$, $B = \frac{1}{C}V^0_1V_1^* + V_2^0V_2^*$, and $C \approx \frac{R_2}{R_1}$.
We can use a binomial approximation on Eqs.~\ref{eq:R1_correction_sol}-\ref{eq:R2_correction_sol} and divide them to get $C$, which lead us to $R_2/R_1 \approx V_2^0V_2^*/V_1^0V_1^*$. We plug this into Eq.~\ref{eq:R2_correction_sol} and get the asymptotic critical time for the learning scale of the second primitive considering the multiplicative effect from the first primitive as following:
\begin{equation}
\label{eq:z_vanilla_final}
    \tau_{\text{vanilla},k=2} = \frac{\sqrt{2\pi}\pi}{\eta_w(T-2)}2^{(T-2)}\frac{1}{(V_1^0V_1^*)^2 + (V_2^0V_{2}^*)^2}.
\end{equation}
\section{Additional Experiments}
\label{appendix:additional}
\begin{figure*}[ht]
\vskip 0.1in
\begin{center}
\centerline{\includegraphics[width=13cm]{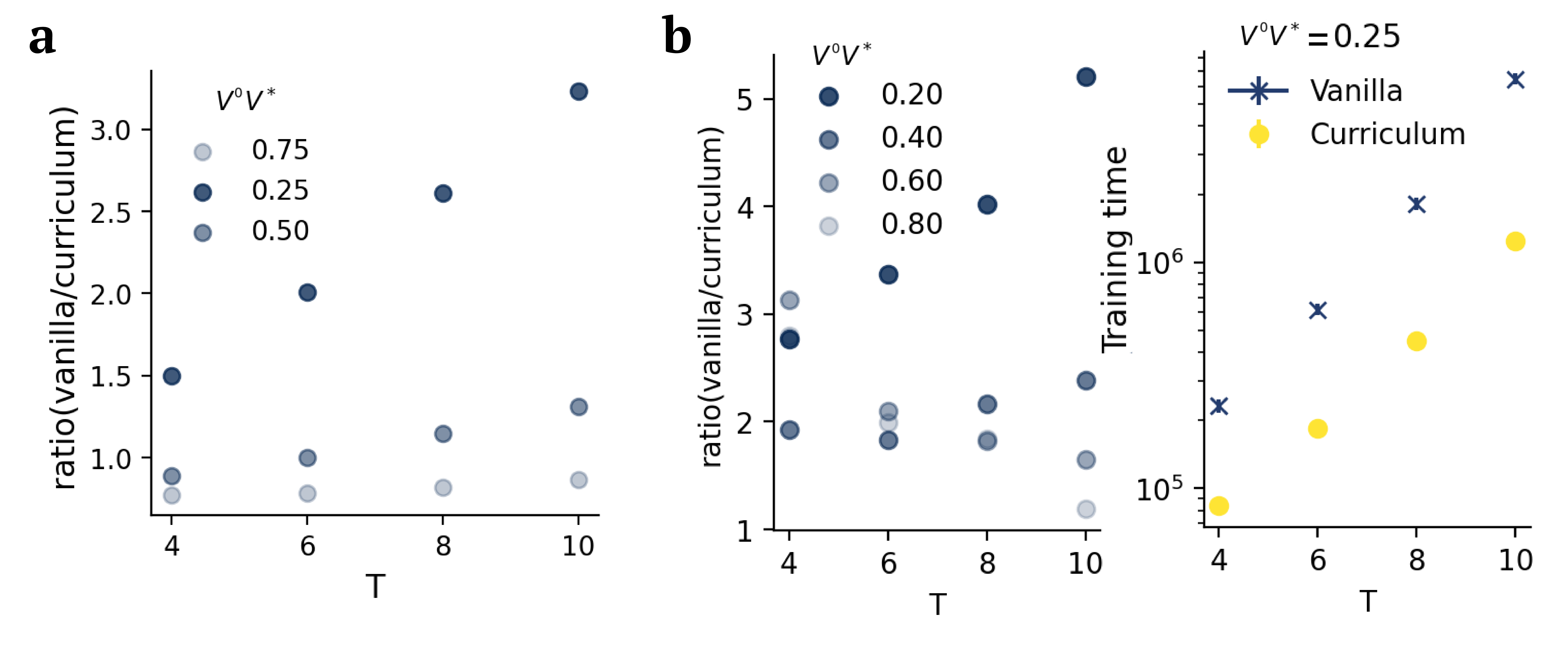}}
\caption{\textbf{a)} The ratio between vanilla learning and curriculum learning in $K=2$ after a hyperparameter search on the learning rates in each of the training protocols. We search over the range [0.01, 10] in logarithmic scale for $\eta_{w, \text{pretrain}}, \eta_{w,\text{composition}}$ and ${\eta_v}$ individually. \textbf{b)} The speed boost effect of the curriculum learning in $K=4$.}
\label{figureA1}
\end{center}
\vskip -0.5in
\end{figure*}

In section~\ref{subsection:learning_speed} and Figure~\ref{figure4}, we showed that for $K=2$ case where the learning rates $\eta_{w, \text{pretrain}}, \eta_{w, \text{compositional}}$ and $\eta_v$ are all fixed to 1. 
Here we extended our experiments with a hyperparameter search on learning rates in each of the training protocols and $K=4$ case to make the curriculum learning effect on learning speed concrete, shown in Figure~\ref{figureA1}.

\begin{figure*}[ht]
\vskip 0.1in
\begin{center}
\centerline{\includegraphics[width=9cm]{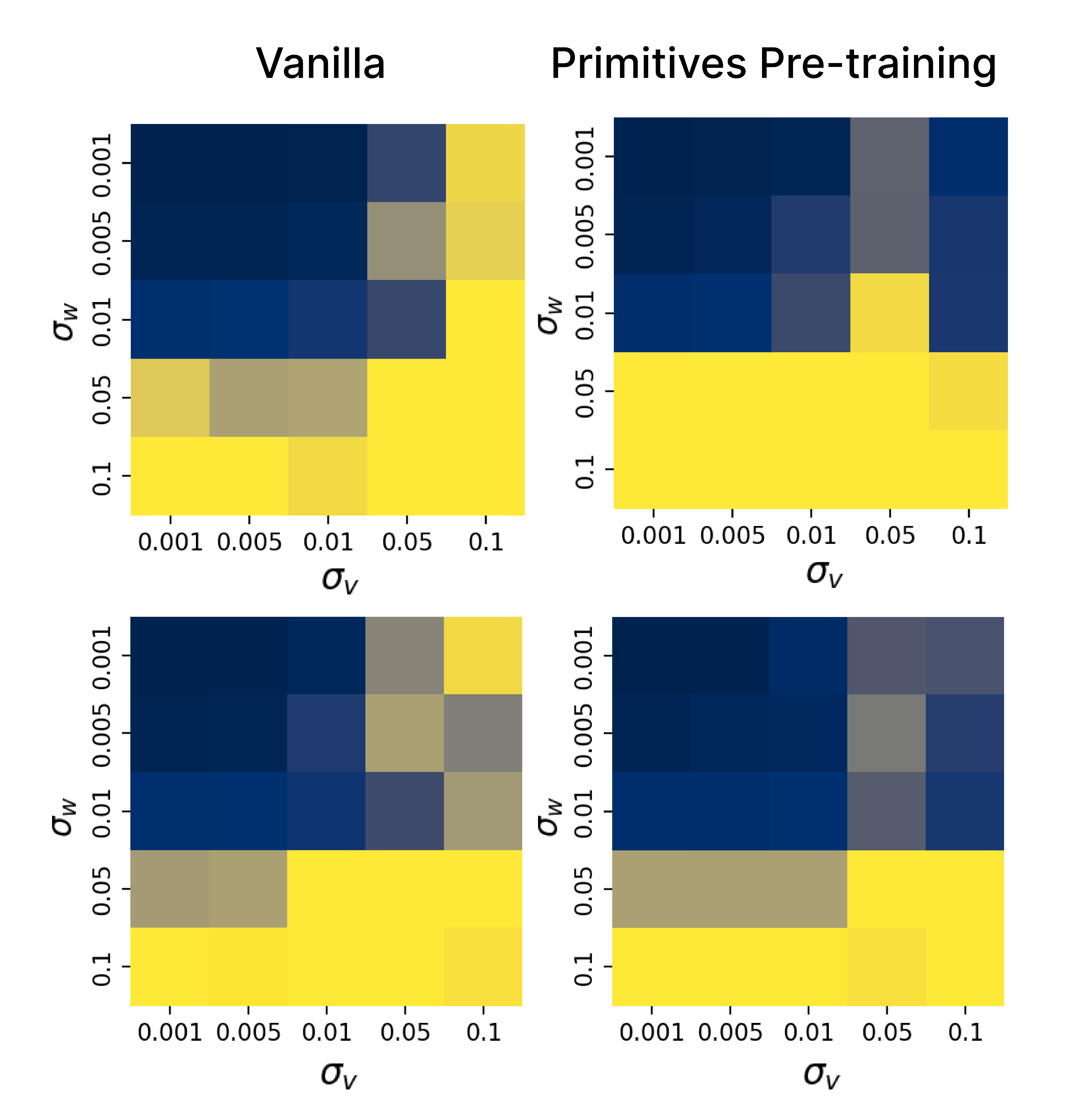}}
\caption{Results from different $V^0V^*$ for Figure. \ref{figure5}b experiment. Top: $K=2$, $V^0V^*=0.5$, Bottom: $K=2$, $V^0V^*=0.75$ }
\label{figureA2}
\end{center}
\vskip -0.5in
\end{figure*}

In addition to Figure~\ref{figure5}b with $V^0V^*=0.25$ case, we show the results for different $V^0V^*$ values, shown in Figure~\ref{figureA2}.

\end{document}